\documentclass[preprint,12pt]{elsarticle}

\usepackage{amssymb}

\newcounter{bla}

\journal{Computer Physics Communications}

\newcommand{\bqa}{\begin{eqnarray}}
\newcommand{\eqa}{\end{eqnarray}}
\newcommand{\nl}{\nonumber \\}
\newcommand{\gloop}{{\tt GLoop}}
\newcommand{\fortran}{{\tt Fortran90}}
\newcommand{\mc}{Monte Carlo}
\newcommand{\avholo}{{\tt OneLOop}}
\newcommand{\qcdloop}{{\tt QCDLoop}}
\newcommand{\eqref}[1]{(\ref{#1})}
\newcommand{\eps}{\epsilon}
\newcommand{\s}{~\,}

\begin{document}

\begin{frontmatter}



\title{\gloop: A {\mc} program to construct higher-loop integrals from lower-loop structures}


\author[a]{Roberto Pittau\corref{author}}

\cortext[author] {Corresponding author.\\\textit{E-mail address:} pittau@ugr.es}
\address[a]{Departamento de F\'isica Te\'orica y del Cosmos, Universidad de Granada, 18071 Granada, Spain}

\begin{abstract}
  We present {\gloop}, a {\fortran} computational framework that allows one to compute by Monte Carlo a certain class of higher-loop integrals in terms of lower-loop building blocks. This is based on a recently introduced method that enables the numerical computation of integrals defined by $i \eps$ deformations acting on single pole singularities without the need for an explicit analytic contour deformation. We provide detailed, worked-out examples and routines to show how our strategy works. These can be used as a starting point for the reader to develop her/his own calculations.






\end{abstract}


\end{frontmatter}



{\bf PROGRAM SUMMARY}
\\\\
\begin{small}
\noindent
{\em Program Title:} \gloop                                       \\
{\em CPC Library link to program files:} (to be added by Technical Editor) \\
{\em Developer's repository link:} {\tt https://www.ugr.es/local/pittau/GLoop/} \\
{\em Code Ocean capsule:} (to be added by Technical Editor)\\
{\em Licensing provisions:} MIT \\
{\em Programming language:} \fortran                                   \\
{\em Nature of problem:} Our aim is to build higher-loop integrals by gluing together lower-loop structures linked by two propagators sharing the same loop momentum.\\
{\em Solution method:} This is achieved by using the {\mc} approach of \cite{Pittau:2021jbs,Pittau:2024ffn} to integrate over the intermediate threshold singularities without performing explicit contour deformations.\\
{\em Additional comments including restrictions and unusual features:} The {\avholo} \cite{vanHameren:2010cp} and {\qcdloop} \cite{Ellis:2007qk} one-loop libraries are used and distributed together with the {\gloop} source code. \\
   \\

\end{small}

\section{Introduction}
The precision of present \cite{CMS,ATLAS} and future \cite{FCC:2025lpp} high-energy particle physics experiments requires a high level of accuracy in the theoretical predictions, which must include ever-increasing orders of radiative corrections.
Two complementary strategies have appeared. Analytic methods based on systems of differential equations \cite{Kotikov:1991pm,Gehrmann:1999as,Henn:2013pwa} have shown their ability to cope with complicated multi-loop calculations
\cite{Caola:2014iua,Gehrmann:2015bfy,Bonciani:2016qxi,Badger:2017jhb,Kudashkin:2017skd,Frellesvig:2019byn,Canko:2020ylt,Agarwal:2021vdh,Abreu:2021asb,Agarwal:2023suw,Abreu:2024fei}. On the other hand, techniques have been developed to deal with the problem in a fully numerical way
\cite{Binoth:2003ak,Bierenbaum:2010cy,Runkel:2019yrs,Capatti:2020xjc,Liu:2022chg,Dubovyk:2022frj,Armadillo:2022ugh,Borinsky:2023jdv,Heinrich:2023til}.

A more direct approach has been presented in \cite{Pittau:2021jbs,Pittau:2024ffn}, where examples have been given on how to produce Monte Carlo (MC) estimates of multi-loop integrals by gluing together lower-loop substructures directly in the Minkowski space.
The key-point of this method is an algorithm that allows one to compute numerically integrals defined via $i \eps$ deformations acting on single-pole singularities \footnote{In multi-loop jargon, they are often referred to as threshold singularities.} without explicitly deforming the integration contour into the
complex plane. 
This techniques requires a non-zero value for $\eps$, but its influence on the result can be lowered close to the machine precision  level, e.g. between $10^{-12}$ and $10^{-9}$ times the largest physical scale appearing in the problem.

In this paper we present {\gloop}, a computational framework written in {\fortran} based on the technique presented in \cite{Pittau:2021jbs,Pittau:2024ffn}.
It is not intended to resolve the issue in its entirety. In contrast, it functions as a reference for the reader who is interested in implementing this method.
As a result, {\gloop} is designed in a highly modular manner, allowing the user to expand the code's capabilities by introducing new cases.

The structure of the paper is as follows. In Section \ref{sec:method} we review the algorithm to integrate numerically over threshold singularities.
Section \ref{gluing} describes our method to glue together lower-loop building blocks into higher-loop integrals. A strategy to improve the MC integration is presented in Section \ref{sec:improving}. Sections \ref{sec:inst} and \ref{sec:gloop} use a three-loop self-energy scalar diagram as an explicit case study
to describe the core structure of {\gloop}. Routines to test the integration algorithm over $\eps$-deformed contours are introduced in Section \ref{sec:testeps}. Finally, Sections \ref{sec:c1}, \ref{sec:d1f} and \ref{sec:d1d} present infrared finite and infrared divergent one-loop integrals as further fully worked-out examples.
\section{Integrating over $\eps$-deformed contours}
\label{sec:method}
Here we briefly outline the method described in \cite{Pittau:2021jbs,Pittau:2024ffn} to integrate over threshold singularities.
Our aim is to flatten the behavior of a $\eps$-deformed singularity parameterized as
\bqa
\label{eq:I0}
I = \int_{-1}^1 dx\, \frac{F(x)}{x+i\eps},
\eqa
where the numerator function $F(x)$ is regular in $x=0$.
To achieve this we introduce a complex integration variable
$
z= \alpha + i \beta
$
related to $x$ by
$
x+i\eps= e^{i \pi (1-z)}.
$
The path in the complex $z$ plan is fixed by requiring $x$ to remain real, which
results in the following relation among $\alpha$ end $\beta$
\bqa
\label{eq:path}
\pi \beta= \ln \frac{\eps}{\sin[\pi (1-\alpha)]},
\eqa
so that
\bqa
x= \frac{\eps}{\tan[\pi (1-\alpha)]}.
\eqa
Using now
\bqa
\label{eq:par1}
dz= d\alpha \left(1+i \frac{d \beta}{d \alpha}\right)=
    d\alpha \left(1+i \frac{x}{\eps}\right)
\eqa
gives
\bqa
\label{eq:I1}
I = -\frac{i \pi}{g_\eps} \int_{{\eps}/{\pi}}^{1-{\eps}/{\pi}}\!\!\!\!d\alpha 
\left( 1+i\frac{x}{\eps} \right) F(x),~~~ g_\eps = 1-\frac{2\eps}{\pi}.
\eqa

Note that the integrand of \eqref{eq:I1} is now regular in $x=0$ for arbitrarily small values of $\eps$. In fact, the $\eps$ dependence is moved to the boundaries of the integration region $x=\pm 1$, far away from the singularity of \eqref{eq:I0}. In practice, this allows one to use tiny numerical values of $\eps$, close to the machine precision level, e.g. between $10^{-12}$ and $10^{-9}$ times the largest physical scale appearing in the problem.
Note also that, if $F(x)$ contains branch cuts in the $x$ complex  plane, the fact that $x$ always lies on the real axis ensures that the right Riemann sheet is automatically taken when $-1 \le x \le 1$. Thus, compared to methods based on contour deformation \cite{Soper:1999xk,Capatti:2019edf,Kermanschah:2021wbk}, one does not have to worry about choosing a path that avoids the pole at $x= -i \eps$ without crossing any cut of $F(x)$.

Equation \eqref{eq:I1} is optimal for integrating over $\alpha$. To flatten
the integration over $\beta$, the parameterization complementary to 
\eqref{eq:par1} is needed, namely
\bqa
\label{eq:par2}
dz= d\beta \left(\frac{d \alpha}{d \beta}+i\right).
\eqa
However, \eqref{eq:path} implies that $\alpha$ is a two-valued function of $\beta$. Therefore, it is necessary to divide \eqref{eq:I1} into two parts
\bqa
\label{eq:Ifb}
I  &=& -\frac{i \pi}{g_\eps}
\int_{{\eps}/{\pi}}^{1/2}\!\! d\alpha \nl
&&\times\left[
 \left(1-i\frac{y_\alpha}{\eps}\right)F(-y_\alpha)
+\left(1+i\frac{y_\alpha}{\eps}\right)F(y_\alpha)
\right], \nl
y_\alpha &=& {\eps}/{\tan(\alpha \pi)},
\eqa
which in terms of $\beta$ produces
\bqa
\label{eq:Ifc}
I &=& - \frac{i \pi}{g_\eps} \int_{\beta_-}^{\beta_+}
d\beta \nl
&& \times \left[
 \left( \frac{\eps}{-y_\beta} +i\right) F(-y_\beta)
-\left( \frac{\eps}{ y_\beta} +i \right)F( y_\beta)
\right],
\eqa
with
\bqa
y_\beta= e^{\pi\beta}\sqrt{1-\left(\frac{\eps}{e^{\pi \beta}}\right)^2},\s\beta_-= \frac{1}{\pi}\ln \frac{\eps}{\sin \eps},\s\beta_+= \frac{\ln \eps}{\pi}.  \nonumber
\eqa

In {\gloop} equations \eqref{eq:Ifb} and \eqref{eq:Ifc} are merged together by means of a multichannel MC approach \cite{Kleiss:1994qy}, so that the complete $1/(x+i \eps)$ behavior of \eqref{eq:I0} is flattened. To increase the robustness of the numerical integration, two additional MC channels are superimposed, namely a flat $x$ distribution and a distribution peaked around $|x|= 1$.   

The described algorithm can be easily extended to any dimensionality. The present version of {\gloop} is able to deal with integrals of the type
\bqa
\int_{-\infty}^\infty \prod_{j=1}^m \left(\frac{d \sigma_j}{\sigma_j \pm i \eps} \right) F(\sigma_1,\sigma_2,\ldots,\sigma_m),
\eqa
with $m$ up to 4.

\section{Gluing substructures}
\label{gluing}
\begin{figure}
\vskip -4.5cm
\hskip -1.2cm
\includegraphics[width=6.5in]{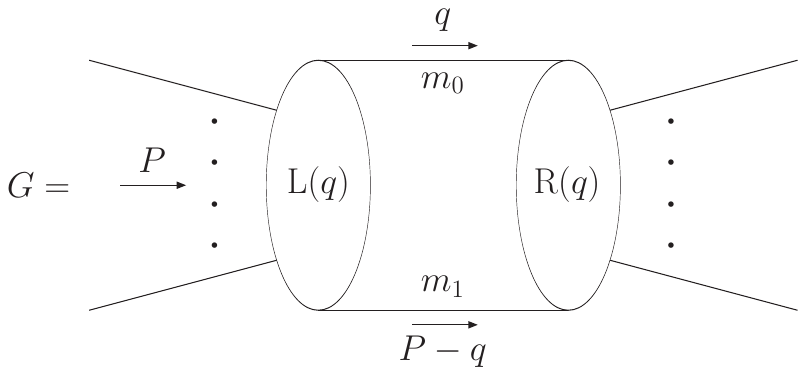}
\vskip -4.35cm
\caption{The diagram $G$ represents the gluing of substructures implemented in {\gloop}.}
\label{fig:1}   
\end{figure}
The type of loop diagrams that can be dealt with {\gloop} is depicted in
Fig.~\ref{fig:1}. Two substructures, represented by the two blobs labelled with L and R, are connected by two propagators with momenta $q$ and $P-q$ and masses $m_0$ and $m_1$, respectively, where $P$ is the sum of all external momenta entering the left blob.
The L and R blobs can contain any number of loops and are computed analytically or semi-numerically as functions of $q$.
We understand that appropriate UV and IR subtractions are implemented in $G$, for instance with the algorithm of \cite{Anastasiou:2018rib}, so that the $q$ integration can be performed in four dimensions. One then has
\bqa
\label{eq:G}
G= \int d^4q\, \frac{{\cal N}(q)}{D_0 D_1},\quad
{\cal N}(q)= {\rm L}(q){\rm R}(q){\rm T(q)},
\eqa
with 
\bqa
D_0= q^2 -m_0^2+i \eps,\quad D_1= (q-P)^2-m_1^2+i \eps, 
\eqa
and where ${\rm T(q)}$ stands for the tensor structures of the two propagators. \footnote{For instance, ${\rm T(q)}= 1$ for scalar propagators.}  
The loop momentum can be parameterized as
\bqa
q^\alpha &=& (q_0,|{\vec q}\,|c_\theta,|\vec q\,|s_\theta s_\phi,|\vec q\,|s_\theta c_\phi ),
\eqa
where
$c_\theta = \cos{\theta}$, $s_\theta = \sin{\theta}$, $c_\phi= \cos\phi$, with
$0 \le \theta \le \pi$ and $0 \le \phi < 2 \pi$.
It is convenient to introduce a mass scale $m$ and work with dimensionless quantities
\bqa
\label{eq:m2}
\begin{tabular}{ll}
  $\omega^\alpha = q^\alpha/m= (t,\rho c_\theta,\rho s_\theta s_\phi,\rho s_\theta c_\phi),$ & $d_i= D_i/m^2= \sigma_i +i \eps,$ \\
  $\tau= P^2/m^2,$ & $\mu_i= m^2_i/m^2,$ 
\end{tabular}
\eqa
in terms of which one has 
\bqa
G = \int d^4\omega \frac{{\cal N}(\omega)}{(\sigma_0+i\eps)(\sigma_1+i \eps)},
\eqa
where
\bqa
\label{eq:d4omega}
\int d^4\omega = \int_{-\infty}^{\infty} \!dt \int_0^{\infty}\!d \rho\rho^2 \int d \Omega =
\int_{-\infty}^{\infty} d\sigma_0 \int_{-\infty}^{\infty} d \sigma_1 \frac{\lambda^{\frac{1}{2}} \Theta(\lambda)}{8 \tau} \int d \Omega.
\eqa
In the last line of \eqref{eq:d4omega} we have performed the change
of variables $(t,\rho) \to (\sigma_0,\sigma_1)$, which gives rise to the appearance of the K\"all\'en function
\bqa
\label{eq:kallen}
\lambda= \lambda(\tau,\sigma_0+\mu_0,\sigma_1+\mu_1).
\eqa
Here we have assumed $\tau > 0$. The case with  $\tau < 0$ can be obtained by analytic continuation \cite{Pittau:2021jbs}. 
By defining
\bqa
\label{eq:N2}
      {\rm N(\sigma_0,\sigma_1)} = \frac{\lambda^{\frac{1}{2}} \Theta(\lambda)}{8 \tau} \int d \Omega\, {\cal N}(\sigma_0,\sigma_1,\Omega)
\eqa
one arrives at
\bqa
\label{eq:G2}
G = \int_{-\infty}^{\infty} \prod_{j=0}^{1} \left(
\frac{d \sigma_j}{\sigma_j+i \eps} \right)
{\rm N(\sigma_0,\sigma_1)}.
\eqa
Equation \eqref{eq:G2} is particularly useful when the integration over the solid angle $\Omega$ in \eqref{eq:N2} can be performed analytically.
In cases when integrating over $c_\theta$ and/or $\phi$ involves threshold singularities to be dealt numerically, alternative representations are
\bqa
\label{eq:G3}
G = \int_{-\infty}^{\infty} \prod_{j=0}^{2} \left(
\frac{d \sigma_j}{\sigma_j+i \eps} \right)
{\rm N(\sigma_0,\sigma_1,\sigma_2)},
\eqa
and
\bqa
\label{eq:G4}
G = \int_{-\infty}^{\infty} \prod_{j=0}^{3} \left(
\frac{d \sigma_j}{\sigma_j+i \eps} \right)
{\rm N(\sigma_0,\sigma_1,\sigma_2,\sigma_3)}.
\eqa
\section{Improving the MC integration}
\label{sec:improving}
In \cite{Pittau:2021jbs} it has been noticed that when the integrands do not
vanish fast enough at large values of $\sigma_{0}$ or $\sigma_{1}$, large cancellations are expected among different integration regions, leading to large MC errors. The procedure to cure this is based on constructing an approximation
${\tilde G}$ of $G$ in \eqref{eq:G2} valid in the asymptotic limit
\bqa
|\sigma_{10}|= |\sigma_1-\sigma_0| \to \infty.
\eqa
Furthermore, ${\tilde G}$ is required to be zero.
In practice, this is obtained by choosing 
\bqa
\label{eq:G2app}
{\tilde G} = \int_{-\infty}^{\infty} d\sigma_{10}\,\Theta[\lambda_0^2(\tau^2+\sigma^2_{10})-\tau^2] \int_{-\infty}^{\infty} d\sigma_0 
\frac{{\rm {\tilde N}(\sigma_0,\sigma_0+\sigma_{10})}}{(\sigma_0+i \eps)(\sigma_0+\sigma_{10}+i \eps)},
\eqa
where the approximated numerator ${\rm {\tilde N}}$ is such that all branch points and poles lie in the lower $\sigma_0$ complex half-plane. Finally,  ${\tilde G}= 0$ is subtracted MC point by MC point from
\eqref{eq:G2}, leading to 
\bqa
G= \int_{-\infty}^{\infty} \prod_{j=0}^{1} \left(
\frac{d \sigma_j}{\sigma_j+i \eps} \right)
\left({\rm N(\sigma_0,\sigma_1)}-
{\rm {\tilde N}(\sigma_0,\sigma_1)}\Theta[\lambda_0^2(\tau^2+\sigma^2_{10})-\tau^2]
\right).
\eqa
The parameter \mbox{$0 \le \lambda_0 \le 1$} corresponds to the variable {\tt alam0} of {\gloop}, whereby when ${\tt alam0}= 0$ (${\tt alam0}= 1$) the subtraction is never (always) performed.

This strategy can be extended to the representations in \eqref{eq:G3} and
\eqref{eq:G4}. For instance, a $\tilde G= 0$ approximation to \eqref{eq:G3} is
\bqa
\label{eq:gtilde}
    {\tilde G} &=& \int_{-\infty}^{\infty} d\sigma_{10}\int_{-\infty}^{\infty} d\sigma_{20}\,{\rm F}_{\rm cut}(\lambda_0,\tau,\sigma_0,\sigma_1,\sigma_2)  \nl
  && \times \int_{-\infty}^{\infty} d\sigma_0
\frac{{\rm {\tilde N}(\sigma_0,\sigma_0+\sigma_{10},\sigma_0+\sigma_{20})}}{(\sigma_0+i \eps)(\sigma_0+\sigma_{10}+i \eps)(\sigma_0+\sigma_{20}+i \eps)},
\eqa
where $\sigma_{20}= \sigma_2-\sigma_0$ and
the approximated numerator ${\rm {\tilde N}}$ is constructed in such a way that all branch points and poles lie in the lower $\sigma_0$ complex half-plane.
Assuming that this approximation gets relevant when
$|\sigma_1-\sigma_0| \to \infty$ or $|\sigma_2-\sigma_0| \to \infty$ determines the cut function 
\bqa
\label{eq:fcut}
      \!\!\!\!\!\!{\rm F}_{\rm cut}(\lambda_0,\tau,\sigma_0,\sigma_1,\sigma_2) &=&
       \Theta[\lambda_0^2(\tau^2+\sigma^2_{10})-\tau^2] \nl
      &&\!\!\!\!+\Theta[\tau^2-\lambda_0^2(\tau^2+\sigma^2_{10})]\,
       \Theta[\lambda_0^2(\tau^2+\sigma^2_{20})-\tau^2].
\eqa
\section{Installing, compiling and running {\gloop}}
\label{sec:inst}
Uncompressing the {\tt GLooop.tar.gz} tarball gives rises to the following structure of files and directories:

\begin{verbatim}
genmodules.f90  introutines    makefile        README
gloop.f90       LICENSE.txt    OneLOop-220331  RESULTS
integrands.f90  loopfunct.f90  QCDLoop-1.98    utlmodules.f90
\end{verbatim}
where the directory {\tt RESULTS} contains the input and output files of the examples presented in this paper.

\vspace{0.3cm}
\noindent To compile {\gloop}
\begin{enumerate}
\item type {\tt make cleanall} to remove old executable, {\tt .o}, {\tt .mod} and
  {\\ .a} files;
\item type {\tt make qcdloop} to create the {\tt libqcdloop.a} library;
\item type {\tt make oneloop} to create the {\tt libavh\_olo.a} library;
\item type {\tt make} to create the {\gloop} executable.
\end{enumerate}
To run {\gloop} type {\tt ./gloop}.
\section{Description of {\gloop}}
\label{sec:gloop}
\begin{figure}[t]
\vskip -4.9cm
\hskip -2.5cm
\includegraphics[width=6.5in]{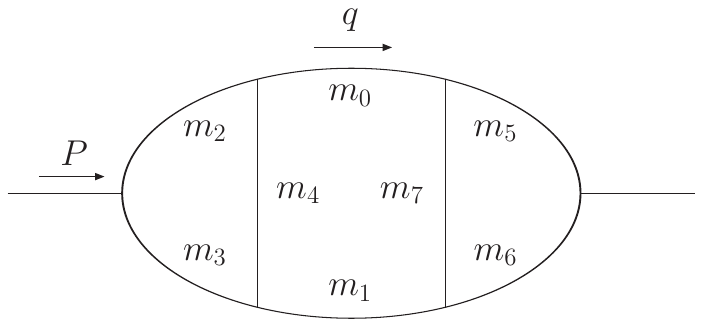}
\vskip -4.55cm
\caption{The three-loop self-energy scalar diagram $\pi^6 S_3(\tau,\{\mu_k\})$ with rescaled invariants $\tau= P^2/m^2$ and $\{\mu_k\} =  \{\mu_0,\mu_1,\mu_2,\mu_3,\mu_4,\mu_5,\mu_6,\mu_7\}=\{m^2_k/m^2\}$.}
\label{fig:2}   
\end{figure}
The main program is {\tt gloop.f90}, which should be modified by the user to load the routines located in the directory {\tt introutines}.
As an illustrative example, we describe here in detail the case of the three-loop self-energy scalar diagram $S_3$ of Fig.~\ref{fig:2} implemented in {\tt introutines/s3.f90}
\bqa
\label{eq:s3}
S_3(\tau,\{\mu_k\})&=& \frac{1}{\pi^6}\int_{-\infty}^{\infty} \prod_{j=0}^{1} \left(
\frac{d \sigma_j}{\sigma_j+i \eps} \right)
\frac{\lambda^{\frac{1}{2}} \Theta(\lambda)}{8 \tau} 
\,(4 \pi)\,C_LC_R, \\ 
\label{eq:s3a}
C_L &=& C(\tau,\sigma_1+\mu_1,\sigma_0+\mu_0,\mu_2,\mu_3,\mu_4), \\
\label{eq:s3b}
C_R &=& C(\tau,\sigma_1+\mu_1,\sigma_0+\mu_0,\mu_5,\mu_6,\mu_7).
\eqa

The file {\tt s3.f90} contains {\tt subroutine gl\_s3inp}, {\tt subroutine gl\_s3out} and {\tt function gl\_s3}. To load them into {\gloop} the first two lines of {\tt gloop.f90} should be set to
\begin{verbatim}
 #define DMNS    2
 #define NAME    gl_s3
\end{verbatim}
The first line defines the dimensionality of the MC integration, while
the second instruction tells {\gloop} to use {\tt function gl\_s3} for the integration and to set the input and output parameters as specified in {\tt subroutine gl\_s3inp} and {\tt subroutine gl\_s3out}, respectively. 
In what follows we describe {\tt subroutine gl\_s3inp}, {\tt subroutine gl\_s3out} and {\tt function gl\_s3}, which, for the reader's convenience, are reported in full in~\ref{app:a},~\ref{app:b} and~\ref{app:c}, and describe the input/output structure of {\tt gloop}. 
\subsection{{\tt subroutine gl\_s3inp}}
The first statement {\tt call checkndim(2)} checks whether the correct value of {\tt DMNS} has been chosen in {\tt gloop.f90}, namely {\tt 2} in the case at hand.
The variable {\tt looptype} sets the one-loop library to be employed. Three options are available, namely {\tt 'oneloop'}, {\tt 'qcdloop'} or {\tt 'notused'}. The last option should be used when no one-loop library is needed.
Finally, the last statements are used to load the input values of the run, namely
$\tau$ and $\mu_k$ $(k= 1\div 7)$.
\subsection{{\tt subroutine gl\_s3out}}
The variable {\tt canal} is there for checking purposes. It represents the analytic value of the integral, when known. If this value is unknown, {\tt canal}  should be set to {\tt c0}, namely the complex zero ${\tt (0.d0,0.d0)}$.
The rest of the routine tells {\gloop} to print out the input values used for the run.
\subsection{{\tt function gl\_s3}}
$S_3$ is constructed here by gluing together the two one-loop triangles of \eqref{eq:s3a} and \eqref{eq:s3b}. As a matter of fact, comparing \eqref{eq:s3} to
\eqref{eq:G}, \eqref{eq:N2} and \eqref{eq:G2} gives $L = C_L$, $R= C_R$ and
$T=1$. The arguments {\tt (y1,by1,y2,by2)} are for internal use only. \footnote{This string of argument is relevant for two-dimensional integrations. In the case of one-, three- or four-dimensional integrations, the arguments should be
{\tt (y1,by1)}, {\tt (y1,by1,y2,by2,y3,by3)} and
{\tt (y1,by1,y2,by2,y3,by3,y4,by4)}, respectively.}
The logical parameters 
{\tt infty0} and {\tt infty1} are set to {\tt .true.} to indicate that the range of the integration variables {\tt s0} and {\tt s1} is between $-\infty$ and
$+\infty$, \footnote{More in general, if {\tt infty{\it n}=.true.} the integration interval of the $n^{th}$ variable {\tt s{\it n}} is
{\mbox [-$\infty$,$\infty$]}, while it is {\mbox [-1,1]} when {\tt infty{\it n}=.false.}.} and the function {\tt smap} takes care of the relevant Jacobian.
{\tt function phinum()} is the numerator structure of \eqref{eq:s3}, where
{\tt kcalc(s0,s1)} evaluates, when positive, the square root of the K\"all\'en function of \eqref{eq:kallen}, whose value is saved in the variable {\tt csklam}.   
The one-loop triangles $C_{L,R}$ are computed analytically by calling {\tt subroutine loop3}. They are multiplied by $i \pi^2$ to compensate for the ${1/(i \pi^2)}$ normalization factor included in {\tt oneloop} and {\tt qcdloop}.
Finally, {\tt denom2(1,1)} is the denominator structure of \eqref{eq:s3}. Its arguments
fix the sign of the $i \eps$ prescription to be used,
${\tt denom2(\pm 1, \pm 1)} = 1/({\tt s0}\pm i \eps)/({\tt s1}\pm i \eps)$.
\subsection{Input/output}
When typing {\tt ./gloop} the code asks for the input values according to the content of {\tt subroutine gl\_s3inp}:
\begin{verbatim}
enter tau,amu0s,amu1s,amu2s,amu3s,amu4s,amu5s,amu6s,amu7s
\end{verbatim}
We use as an illustrative example 10,1,1,2,3,4,5,6,7.
After entering this string, {\gloop} asks for the number of MC points, {\tt npoints}.
Setting, for instance, {\tt npoints= 1000000} points, produces four iterations of $10^6$ points each, \footnote{The number of iterations is controlled by the argument of {\tt call setnrtpmax} in {\tt gloop.f90}.} whose results are averaged in the final output
\begin{verbatim}
 ---------------------averaged result------------------------  
   
 jseed(1)= 93253   jseed(2)= 42930   looptype= oneloop
   
 m^2= 0.1000D+01  eps= 0.1000D-08 
   
 real=  0.114668D+00  +-  0.118030D-02 | analr= not available
 imag= -0.416500D-01  +-  0.191633D-02 | anali= not available
   
 ------------------------------------------------------------  
\end{verbatim}
where {\tt real} and {\tt imag} are the computed real and imaginary parts of the MC integration, with the corresponding MC errors, to be compared, when available, to the analytic results {\tt analr} and {\tt anali}. Finally, {\tt jseed(1)} and {\tt jseed(2)} are the seeds of the random number generator used for the MC integration, {\tt looptype} specifies the one-loop library used in the run, \verb|m^2| and ${\tt eps}$ are the mass scale $m^2$ and the value of $\eps$ of \eqref{eq:m2}, respectively. \footnote{The default value of $\eps$ is $10^{-9}$ and can be changed by modifying the variable {\tt epsval} in {\tt gloop.f90}.}
The time to produce this output on a single 2.2 GHz processor is
of about 95 s.
\section{Testing the integrations over $\eps$-deformed contours}
\label{sec:testeps}
In this section we describe four routines whose main scope is to illustrate the performance of the method presented in section~\ref{sec:method}. They also serve as simple examples for the user to familiarize her/himself with the structure of {\gloop}.
\subsection{The one-dimensional integral {\tt gl\_test1}}
\label{sec:I1}
This routine computes numerically the integral
\bqa
\label{eq:1}
I_1(\tau) = \int_{-1}^1 d \sigma \frac{\ln(\sigma+\tau-i\eps)}{\sigma+i\eps}.
\eqa
It is located in {\tt introutines/test1.f90} and can be loaded by setting the
first two lines of {\tt gloop.f90} to
\begin{verbatim}
 #define DMNS    1
 #define NAME    gl_test1
\end{verbatim}
Note that when $-1 < \tau < 1$ the branch point of the logarithm is moved away from the integration contour by the $-i \eps$. This represents the simplest check on the performance of our numerical algorithm: $\sigma$ is always kept real in the interval $-1 \le \sigma \le 1$, so that the right branch of the logarithm is automatically taken.  
\begin{table}[t]
\caption{
  Numerical estimates of $I_1(\tau)$ in \eqref{eq:1} for several values of $\tau$.
The analytic result is reported in \eqref{eq:2}. Numbers obtained with $4 \times 10^7 $ MC shots and $\eps= 10^{-9}$. MC errors between parentheses. The time to produce each entry in the table on a single 2.2 GHz processor is of about 8.6 s.}
\label{tab:1}       
\begin{tabular}{rll}  
\hline\noalign{\smallskip}
 $\tau$   & MC result & Analytic result \\
\noalign{\smallskip}\hline\noalign{\smallskip}
    -1.5  & -1.1283(1)$\times 10^{1}$ $-$$i$ 1.2739(1)& -1.1283$\times 10^{1}$ $-$$i$ 1.2738\\ 
    -0.5  & -1.3777(2)$\times 10^{1}$ $+$$i$ 4.3563(8) & -1.3774$\times 10^{1}$ $+$$i$ 4.3552\\ 
     0.5  &  3.9043(7)                $+$$i$ 4.3554(8) &  3.9042               $+$$i$ 4.3552\\ 
     1.5  &  1.4131(2)                $-$$i$ 1.2741(2) &   1.4130              $-$$i$ 1.2738 \\ 
\noalign{\smallskip}\hline
\end{tabular}
\end{table}

In Table \ref{tab:1} we compare the numerical results obtained with {\gloop} to the analytic value
\bqa
\label{eq:2}
I_1^{\rm ana}(\tau)= {\rm Li}_2\left(\frac{1}{\tau -i \eps}\right)-{\rm Li}_2\left(-\frac{1}{\tau -i \eps}\right)
- i \pi \ln(\tau -i \eps).
\eqa

\subsection{The two-dimensional integral {\tt gl\_test2}}
\label{sec:I2}
This routine computes numerically the integral
\bqa
\label{eq:1}
I_2 = \int_{-1}^1 d \sigma_1 \int_{-\infty}^\infty d \sigma_2
\frac{{\rm N}(\sigma_1,\sigma_2)}{(\sigma_1+i\eps)(\sigma_2-i\eps)},
\eqa
where
\bqa
    {\rm N}(\sigma_1,\sigma_2)= \frac{\sigma_1}{(\sigma_2^2+1)} \ln(\sigma_1+\sigma_2-i \eps).
\eqa
It is located in {\tt introutines/test2.f90} and can be loaded by setting the
first two lines of {\tt gloop.f90} to
\begin{verbatim}
 #define DMNS    2
 #define NAME    gl_test2
\end{verbatim}

With $4 \times 10^7 $ MC shots and $\eps= 10^{-9}$ {\gloop} gives
\bqa
\label{eq:3}
I_2= 9.870(2) + i\, 8.24(5) \times 10^{-1}
\eqa
to be compared to the analytic value
\bqa
\label{eq:4}
I_2^{\rm ana}= i \pi \left(\ln 2-2+\frac{\pi}{2}\right)+\pi^2
= 9.86960 + i\, 8.29203 \times 10^{-1}.
\eqa
The time to produce the result of \eqref{eq:3} is of about 22 s on a single 2.2 GHz processor.
\subsection{The three-dimensional integral {\tt gl\_test3}}
\label{sec:I3}
This routine computes numerically the integral
\bqa
\label{eq:5}
I_3 = \int_{-\infty}^\infty d \sigma_1
      \int_{-\infty}^\infty d \sigma_2
      \int_{-\infty}^\infty d \sigma_3
\frac{{\rm N}(\sigma_1,\sigma_2,\sigma_3)}{(\sigma_1+i\eps)(\sigma_2+i\eps)(\sigma_3-i\eps)},
\eqa
with
\bqa
    {\rm N}(\sigma_1,\sigma_2,\sigma_3)= {\rm sgn(\sigma_1)}
\Theta(|\sigma_1|-|\sigma_2|)
    \frac{\sigma_2}{(\sigma_1^2+2)(\sigma_3^2+1)} \ln(\sigma_2+\sigma_3-i \eps).
\eqa
It is located in {\tt introutines/test3.f90} and can be loaded by setting the
first two lines of {\tt gloop.f90} to
\begin{verbatim}
 #define DMNS    3
 #define NAME    gl_test3
\end{verbatim}

With $4 \times 10^7 $ MC shots and $\eps= 10^{-9}$ {\gloop} gives
\bqa
\label{eq:6}
I_3= 2.191(2) \times 10^1 +i\,7.00(3)
\eqa
to be compared to the analytic value
\bqa
\label{eq:7}
I_3^{\rm ana} &=& \frac{\pi^3}{\sqrt{2}}+ i \pi^2
\left(
(1+\sqrt{2}) \ln(1+\sqrt{2})-\sqrt{2}
\right) \nl
&=& 2.19247 \times 10^1 +i\,7.04305.
\eqa
The time to produce the result of \eqref{eq:6} is of about 27 s on a single 2.2 GHz processor.
\subsection{The four-dimensional integral {\tt gl\_test4}}
\label{sec:I4}
This routine computes numerically the integral
\bqa
\label{eq:8}
I_4 = \int_{-\infty}^\infty \!d \sigma_1
      \int_{-\infty}^\infty \!d \sigma_2
      \int_{-1}^1 \!d \sigma_3
      \int_{-1}^1 \!d \sigma_4
\frac{{\rm N}(\sigma_1,\sigma_2,\sigma_3,\sigma_4)}{(\sigma_1+i\eps)(\sigma_2-i\eps)(\sigma_3+i\eps)(\sigma_4+i\eps)},
\eqa
where
\bqa
    {\rm N}(\sigma_1,\sigma_2,\sigma_3,\sigma_4) &=&
\Theta(\sigma_1+3)
\Theta(5-\sigma_1)
\Theta(\sigma_2+2)\Theta(4-\sigma_2) \nl
&&\times (1+\sigma_3)(1-\sigma_4). 
\eqa
It is located in {\tt introutines/test4.f90} and can be loaded by setting the
first two lines of {\tt gloop.f90} to
\begin{verbatim}
 #define DMNS    4
 #define NAME    gl_test4
\end{verbatim}

With $4 \times 10^7 $ MC shots and $\eps= 10^{-9}$ {\gloop} gives
\bqa
\label{eq:9}
I_4= -1.412(4) \times 10^2 + i\,8.3(4)
\eqa
to be compared to the analytic value
\bqa
\label{eq:7}
I_4^{\rm ana} &=& -(\pi^2+4)\left(\ln\frac{5}{3}\ln 2+\pi^2+i \pi \ln \frac{5}{6} \right) \nl
&=& -1.41798 \times 10^2 + i\,7.94423.
\eqa
The time to produce the result of \eqref{eq:9} is of about 33 s on a single 2.2 GHz processor.
\section{The routine {\tt gl\_c1loop}}
\label{sec:c1}
\begin{figure}[t]
\vskip -4.5cm
\hskip -1.8cm
\includegraphics[width=6.5in]{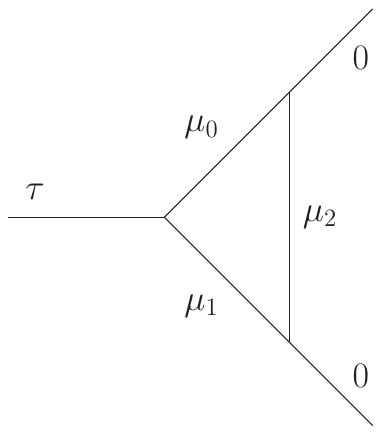}
\vskip -4.25cm
\caption{The rescaled one-loop triangle of \eqref{eq:c1loop}. The two external lines on the right are massless, $\tau= P^2/m^2 > 0$ and $\mu_i= m^2_i/m^2$.}
\label{fig:3}   
\end{figure}
This routine computes numerically the rescaled one-loop triangle of Fig.~\ref{fig:3}
\bqa
\label{eq:c1loop}
C(\tau,0,0,\mu_0,\mu_1,\mu_2)&=&
\int d^4\omega \frac{1}{(\sigma_0+i\eps)(\sigma_1+i \eps)(\sigma_2+i \eps)}, \nl
\sigma_2 &=& \frac{1}{2}\left(A_2
-c_\theta \lambda^{\frac{1}{2}}\right),\nl
A_2 &=& \sigma_0+\sigma_1 +\mu_0 +\mu_1-2\mu_2-\tau.
\eqa
It is located in {\tt introutines/c1loop.f90} and can be loaded by setting the
first two lines of {\tt gloop.f90} to
\begin{verbatim}
 #define DMNS    2
 #define NAME    gl_c1loop
\end{verbatim}

Using \eqref{eq:d4omega} and integrating over $c_\theta$ gives
\bqa
C(\tau,0,0,\mu_0,\mu_1,\mu_2) &=& \int_{-\infty}^{\infty} \prod_{j=0}^{1} \left(
\frac{d \sigma_j}{\sigma_j+i \eps} \right)
     {\rm N(\sigma_0,\sigma_1)}, \\
\label{eq:NC}     
{\rm N(\sigma_0,\sigma_1)} &=& \frac{\pi}{2 \tau} \Theta(\lambda)\ln \frac{A_2+i \eps +\lambda^{\frac{1}{2}}}{A_2+i \eps -\lambda^{\frac{1}{2}}}.    
\eqa
As described in Section \ref{sec:improving}, given that
$
\lim_{\sigma_0,\sigma_1\to \infty}{\rm N(\sigma_0,\sigma_1)} \sim {\rm constant},
$
we need to construct a $|\sigma_{10}| \to \infty$ approximation ${\rm \tilde N(\sigma_0,\sigma_1)}$ of ${\rm N(\sigma_0,\sigma_1)}$ such that
$
\int_{-\infty}^{\infty} \prod_{j=0}^{1} \left(
\frac{d \sigma_j}{\sigma_j+i \eps} \right)
     {\rm \tilde N(\sigma_0,\sigma_1)}= 0.
     $
     ${\rm \tilde N(\sigma_0,\sigma_1)}$ is obtained by replacing $\lambda \to \tilde \lambda$ and $A_2 \to \tilde A_2$ in \eqref{eq:NC}, where
\bqa
\label{eq:appr}     
\tilde \lambda &=& (\sigma_1-\sigma_0+\mu_1-\mu_0)^2-4\tau \mu_2, \nl
\tilde A_2 &=& \sigma_0+\sigma_1+\mu_0+\mu_1-2\mu_2. 
\eqa
With this choice $A_2^2-{\lambda}^2= {\tilde A_2}^2-{\tilde \lambda}^2$,
so that small values of the argument of the logarithm  in \eqref{eq:NC} are always well approximated.
After performing this local subtraction one arrives at
\bqa
&&C(\tau,0,0,\mu_0,\mu_1,\mu_2) = \int_{-\infty}^{\infty} \prod_{j=0}^{1} \left(
\frac{d \sigma_j}{\sigma_j+i \eps} \right) \nl
&&~~~~~\times\left[{\rm N(\sigma_0,\sigma_1)}-
  \Theta[\lambda_0^2(\tau^2+(\sigma_1-\sigma_0)^2)-\tau^2]
  {\rm \tilde N(\sigma_0,\sigma_1)}\right]. 
\eqa

With $4 \times 10^7 $ MC shots, $\eps= 10^{-9},~\tau= 10,~\mu_0= 1,~\mu_1=2,~\mu_2= 3$ and $\lambda_0= 0.7$ {\gloop} produces
\bqa
C(\tau,0,0,\mu_0,\mu_1,\mu_2) = 3.345(6) -i\, 2.988(4),
\eqa
to be compared to the analytic value
\bqa
C^{\rm ana}(\tau,0,0,\mu_0,\mu_1,\mu_2)= 3.34507 -i\, 2.98595.
\eqa
The time to produce this output on a single 2.2 GHz processor is
of about 23s.

\section{The routine {\tt gl\_d1floop}}
\label{sec:d1f}
\begin{figure}[t]
\vskip -4.5cm
\hskip -1.8cm
\includegraphics[width=6.5in]{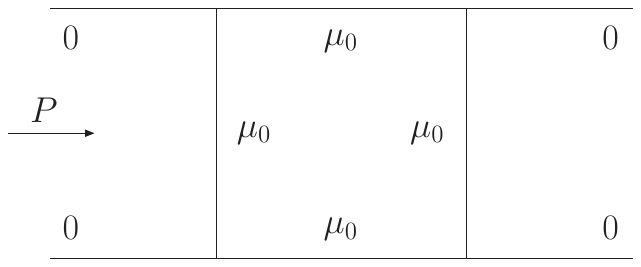}
\vskip -4.25cm
\caption{The infrared finite rescaled one-loop box of \eqref{eq:d1floop} with all equal internal masses $\mu_0= m^2/P^2$. The external lines are massless.}
\label{fig:4}   
\end{figure}

This routine computes numerically the rescaled one-loop box of Fig.~\ref{fig:4}
\bqa
\label{eq:d1floop}
D(x,\mu_0)&=& \int d^4\omega \frac{1}{(\sigma_0+i\eps)(\sigma_1+i\eps)(\sigma_2+i\eps)(\sigma_3+i\eps)}, \nl
\sigma_0&=& t^2-\rho^2-\mu_0,~~~~~\sigma_1~=~ \sigma_0+1-2t, \nl
\sigma_2&=& \sigma_0-t+\rho c_\theta,~~~~~\sigma_3~=~ a+b s_\phi,
\eqa
where
\bqa
\label{eq:ab}
a &=& \sigma_0+(\sigma_2-\sigma_0)(1-2x)+x(\sigma_1-\sigma_0-1), \nl
b &=& 2 \sqrt{x(1-x)}\sqrt{(\sigma_2-\sigma_0)(\sigma_1-\sigma_0-1-(\sigma_2-\sigma_0))-\mu_0-\sigma_0}.
\eqa
The variable $x$ is minus the ratio of the $t_{inv}$ and $s_{inv}$ Mandelstam variables
$
x= -{t_{inv}}/{s_{inv}}
$,
so that  $0 \le x \le 1$ in the physical region.
$D(x,\mu_0)$ is located in {\tt introutines/d1floop.f90} and can be loaded by setting the
first two lines of {\tt gloop.f90} to
\begin{verbatim}
 #define DMNS    3
 #define NAME    gl_d1floop
\end{verbatim}

Using \eqref{eq:d4omega} and integrating over $\phi$ allows one to recast
\eqref{eq:d1floop} as in \eqref{eq:G3}
\bqa
D(x,\mu_0) = \int_{-\infty}^{\infty} \prod_{j=0}^{2} \left(
\frac{d \sigma_j}{\sigma_j+i \eps} \right)
{\rm N(\sigma_0,\sigma_1,\sigma_2)},
\eqa
with
\bqa
{\rm N(\sigma_0,\sigma_1,\sigma_2)}= \frac{\pi}{2}\Theta(\lambda)
\Theta(\sigma_2^+-\sigma_2)
\Theta(\sigma_2-\sigma_2^-)\, {\rm I_\phi}(a,b^2),
\eqa
where
\bqa
\label{eq:lamsig}
\lambda = \lambda(1,\sigma_0+\mu_0,\sigma_1+\mu_0),~~
\sigma_2^\pm = (\sigma_1+\sigma_0-1 \pm \lambda^{\frac{1}{2}})/2
\eqa
and
\bqa
\label{eq:Iphi}
{\rm I_\phi}(a,b^2) &=&
 \frac{1}{2 \pi}\int_{-\pi}^0  \frac{d \phi}{a^+  +s_\phi b^- }
 +\frac{1}{2 \pi}\int_{0}^{\pi} \frac{d \phi}{a^+ +s_\phi b^+ } \nl
 &=& \frac{1}{2 \pi}
 \sum_{\alpha= \pm}~
 \frac{1}{a^+ \sqrt{1-(\frac{b^\alpha}{a^+})^2}}
       \left(\pi-2\alpha \arcsin\Bigl(\frac{b^\alpha}{a^+}\Bigr)\right),
 \eqa
with $a^+= a + i \eps$ and $b^\pm = \sqrt{b^2 \pm i \eps}$.
Note that \eqref{eq:Iphi} is written in a form suitable for analytic continuation into the unphysical regions $x < 0$ and $x > 1$.

With $4 \times 10^8$ MC shots, $\eps= 10^{-9}$, $x= 0.7$ and $\mu_0= 0.15$
{\gloop} produces
\bqa
D(0.7,0.15)= -1.005(4) \times 10^{2} + i\, 3.70(4) \times 10^{1}
\eqa
to be compared to the analytic value
\bqa
D^{\rm ana}(0.7,0.15)=  -1.00943 \times 10^{2} + i\, 3.69428 \times 10^{1}.
\eqa
The time to produce this output on a single 2.2 GHz processor is
of about 6.5 min.

\section{The routine {\tt gl\_d1dloop}}
\begin{figure}[t]
\vskip -4.5cm
\hskip -0.08cm
\includegraphics[width=5.5in]{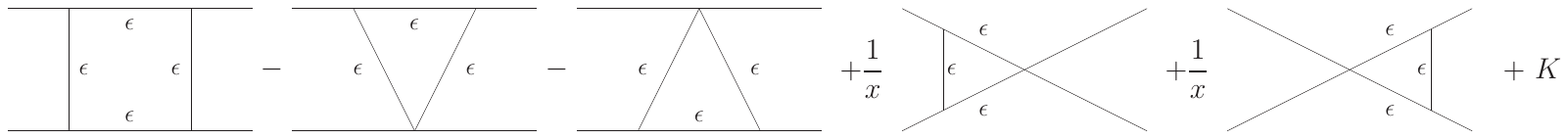}
\vskip -4.25cm
\caption{The subtracted rescaled one-loop box of \eqref{eq:d1dloop} with all equal internal masses $\eps$. The external lines are massless and $x= -{t_{inv}}/{s_{inv}}$. Choosing $K$ as in \eqref{eq:K} cancels the finite part of the triangles, so that, when $\eps \to 0$, equation \eqref{eq:d1dloop} reproduces the finite part of the infrared divergent box.}
\label{fig:5}   
\end{figure}

In this section we show how to deal with divergent integrals. Our starting point is a four-dimension representation where the appropriate IR and/or UV subtractions are implemented as described in \cite{Anastasiou:2018rib}.
As an example we consider the rescaled one-loop box of Fig.~\ref{fig:5}, that develops IR divergences when all the internal and external masses are zero. Such infinities are subtracted by adding the contribution of the four three-point diagrams in the figure, while the constant $K$ is such that it compensates the finite part of the triangles. Since this combination of integrals is IR finite, one is free to compute it with a small value $\eps \to 0$ of the internal masses. We choose for $\eps$ the same numerical value used to deal with threshold singularities, so that its influence remains at the same level of the error induced by the method described in Section \ref{sec:method}, namely close to the machine precision level.

Putting all the terms together gives
\bqa
\label{eq:d1dloop}
D_{\rm fin}(x) = \int_{-\infty}^{\infty} \prod_{j=0}^{2} \left(
\frac{d \sigma_j}{\sigma_j+i \eps} \right)
{\rm N_{fin}(\sigma_0,\sigma_1,\sigma_2)},
\eqa
where \footnote{The normalization factor of the first term of \eqref{eq:Nfin} is explained by the fact that weights computed at $\sigma_j$ and $-\sigma_j$ are always considered together in {\gloop}, so that $\int_{-\infty}^{\infty} 
\frac{d \sigma_j}{\sigma_j+i \eps}= \lim_{\Lambda \to \infty} \int_{-\Lambda}^{\Lambda}
\frac{d \sigma_j}{\sigma_j+i \eps}= -i \pi$.}
\bqa
\label{eq:Nfin}
&&\!\!\!\!\!\!\!\!\!{\rm N_{fin}(\sigma_0,\sigma_1,\sigma_2)}= K/(-i\pi)^3\nl 
&&\!\!\!\!\!\!\!+\frac{\pi}{2}\Big[\Theta(\lambda)
\Theta(\sigma_2^+-\sigma_2)
\Theta(\sigma_2-\sigma_2^-)
\left[
{\rm I_\phi}(a,b^2)\left(1-\sigma_0-\sigma_1+\sigma_2/x \right) +1/x
\right]\Big]_{\mu= \eps} \nl
\eqa
with $a$, $b^2$, $\lambda$ and $\sigma_2^\pm$ given in \eqref{eq:ab} and
\eqref{eq:lamsig}, and
\bqa
\label{eq:K}
K= \frac{i \pi^2}{x}(\pi^2-\ln^2x).
\eqa

\label{sec:d1d}

When $|\sigma_1-\sigma_0| \to \infty$ or $|\sigma_2-\sigma_0| \to \infty$ \eqref{eq:d1dloop} is plagued by large cancellations among different integration regions. Thus, we apply the strategy of Section \ref{sec:improving} and construct an asymptotic approximation ${\rm {\tilde N_{fin}}(\sigma_0,\sigma_1,\sigma_2)}$ of \eqref{eq:Nfin} such that
\bqa
\int_{-\infty}^{\infty} \prod_{j=0}^{2} \left(
\frac{d \sigma_j}{\sigma_j+i \eps} \right)
{\rm {\tilde N_{fin}}(\sigma_0,\sigma_1,\sigma_2)}= 0.
\eqa
${\rm {\tilde N_{fin}}}$ is constructed from the second line of \eqref{eq:Nfin}
by replacing  
\bqa
\lambda &\to& {\tilde \lambda}= (\sigma_1-\sigma_0)^2-4 \mu_0 \nl
\sigma_2^\pm &\to& {\tilde \sigma}_2^\pm= (\sigma_1+\sigma_0 \pm {\tilde \lambda}^{\frac{1}{2}})/2 \nl
a &\to& {\tilde a}= \sigma_0+(\sigma_2-\sigma_0+x)(1-2x)+x(\sigma_1-\sigma_0)  \nl
b^2 &\to& {\tilde b}^2= 4x(1-x)[(\sigma_2-\sigma_0)(\sigma_1-\sigma_2)-\mu_0-x(x+2\sigma_2-\sigma_1-\sigma_0)], \nl
\eqa
where ${\tilde \lambda}$ and ${\tilde \sigma}_2^\pm$ are chosen with the same logic of \eqref{eq:appr} and ${\tilde a}$, ${\tilde b}^2$ are such that
$a^2-b^2= {\tilde a}^2-{\tilde b}^2$.
This produces
\bqa
\label{eq:d1dloopf}
D_{\rm fin}(x) &=& \int_{-\infty}^{\infty} \prod_{j=0}^{2} \left(
\frac{d \sigma_j}{\sigma_j+i \eps} \right) \nl
&&\times \left(
     {\rm N_{fin}(\sigma_0,\sigma_1,\sigma_2)}
-{\rm F}_{\rm cut}(\lambda_0,1,\sigma_0,\sigma_1,\sigma_2)
     {\rm {\tilde N_{fin}}(\sigma_0,\sigma_1,\sigma_2)}\right)
\eqa
where ${\rm F}_{\rm cut}$ is given in \eqref{eq:fcut}.

\noindent Equation \eqref{eq:d1dloopf} is implemented in {\tt introutines/d1dloop.f90} and can be loaded by setting the first two lines of {\tt gloop.f90} to
\begin{verbatim}
 #define DMNS    3
 #define NAME    gl_d1dloop
\end{verbatim}

To carefully check all the local cancellations of \eqref{eq:d1dloopf} we performed a high-statistics run. With $32 \times 10^9$ MC shots
\footnote{To produce 32 iterations one has to replace in {\tt gloop.f90} the statement {\tt call setnrtpmax(4)} with {\tt call setnrtpmax(32)}.}, $\eps= 10^{-9}$, $x= 0.3$ and $\lambda_0= 0.5$  {\gloop} produces
\bqa
D_{\rm fin}(0.3)= 2.489(1) \times 10^{2} + i\, 3.246(1) \times 10^{2}
\eqa
to be compared to the analytic value
\bqa
D_{\rm fin}^{\rm ana}(0.3) &=& \left.\frac{i \pi^2}{x}(\pi^2+2i\pi \ln x)\right|_{x= 0.3} \nl
&=&2.48871 \times 10^{2} + i\, 3.24697 \times 10^{2}.
\eqa
The time to produce this output on a single 2.2 GHz processor is
of about 17 hours.

\section{Conclusion and outlook}
\begin{figure}[t]
\vskip -4.9cm
\hskip -2.5cm
\includegraphics[width=6.5in]{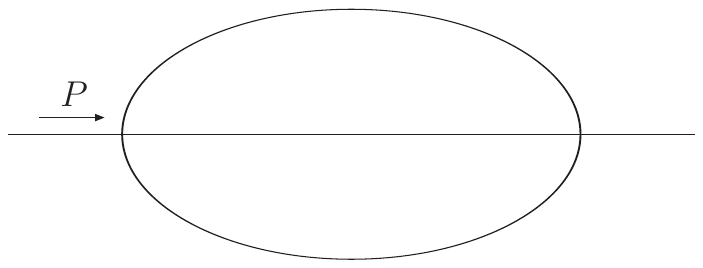}
\vskip -4.55cm
\caption{An example of a diagram that {\gloop} cannot handle in its current form.}
\label{fig:6}   
\end{figure}
We have released the {\fortran} MC code {\gloop} that can be used to construct higher-loop integrals by gluing together lower-loop building blocks. It is a highly modular computational framework based on the algorithm of \cite{Pittau:2021jbs,Pittau:2024ffn} that allows one to perform an accurate numerical integration over threshold singularities.
We have presented a few fully worked-out examples that serve as guidelines for the users willing to implement their own calculations.

The present version of the code is able to glue sub-structures linked by two propagators only. An example of diagram that cannot be dealt with {\gloop} is given in Fig.~\ref{fig:6}. While it is feasible, adding structures connected by more than two propagators necessitates substantial code changes. We reserve this matter for future investigations.

\section*{Acknowledgment}
This work has been partially supported by the Spanish Ministry of Science
and Innovation and SRA (10.13039/501100011033), with ERDF under the grant PID2022-139466NB-C22 and by the Junta de Andaluc\'ia grant FQM 101.

\appendix

\section{The {\tt subroutine gl\_s3inp}}
\label{app:a}
\begin{verbatim}
  subroutine gl_s3inp
   use kinvar
   use generationvariables
   use loopfunctions
   implicit none
   call checkndim(2)
   looptype='oneloop'
   write(*,*) 'enter tau,amu0s,amu1s,amu2s,amu3s,amu4s,amu5s,amu6s,amu7s'
   read*,tau,amu0s,amu1s,amu2s,amu3s,amu4s,amu5s,amu6s,amu7s
  end subroutine gl_s3inp
\end{verbatim}

\section{The {\tt subroutine gl\_s3out}}
\label{app:b}
\begin{verbatim}
  subroutine gl_s3out
   use cvalues
   use kinvar
   implicit none
   canal= c0    
   write(*,*) '      '
   write(*,*) 'The integrand is gl_s3'
   write(*,*) '      '
   write(*,*) 'tau  =',tau
   write(*,*) 'amu0s=',amu0s
   write(*,*) 'amu1s=',amu1s
   write(*,*) 'amu2s=',amu2s
   write(*,*) 'amu3s=',amu3s
   write(*,*) 'amu4s=',amu4s
   write(*,*) 'amu5s=',amu5s
   write(*,*) 'amu6s=',amu6s
   write(*,*) 'amu7s=',amu7s
  end subroutine gl_s3out 
\end{verbatim}

\section{The {\tt function gl\_s3}}
\label{app:c}
\begin{verbatim}
 function gl_s3(y1,by1,y2,by2)
   ! 
   ! Dimensionless 3-loop self-energy
   ! (S_3 of 2110.12885 [hep-ph])
   !
   use epsvalue
   use cvalues
   use pivalue
   use kinvar
   use generationvariables
   implicit none
   complex(kind(1.d0)) :: gl_s3
   real(kind(1.d0)), intent(in) :: y1,by1,y2,by2
   real(kind(1.d0)) :: s0,s1
   logical, parameter :: infty0=.true.,infty1=.true.
   call storeinf(infty0,infty1)   
   s0= smap(infty0,y1,by1)
   s1= smap(infty1,y2,by2)
   gl_s3= phinum()/denom2(1,1)
   contains
!
#include "denominators/denom2.f90"
   function phinum()
     use functionals
     use loopfunctions
     implicit none
     complex(kind(1.d0)) :: phinum
     complex(kind(1.d0)) :: ores(0:2),carg
     phinum= c0
     call kcalc(s0,s1); if(.not.positive) return
     call loop3(ores,tau,s11,s00,amu2s,amu3s,amu4s)
     carg= ores(0)*ci*pi*pi
     call loop3(ores,tau,s11,s00,amu5s,amu6s,amu7s)
     carg= carg*ores(0)*ci*pi*pi
     phinum= (csklam/8.d0/tau)*(4.d0*pi)*(carg)/pi**6
   end function phinum
 end function gl_s3
\end{verbatim}



\bibliographystyle{elsarticle-num}
\bibliography{gloop.bib}

@article{Pittau:2021jbs,
    author = "Pittau, Roberto and Webber, Bryan",
    title = "{Direct numerical evaluation of multi-loop integrals without contour deformation}",
    eprint = "2110.12885",
    archivePrefix = "arXiv",
    primaryClass = "hep-ph",
    doi = "10.1140/epjc/s10052-022-10008-6",
    journal = "Eur. Phys. J. C",
    volume = "82",
    number = "1",
    pages = "55",
    year = "2022"
}

@article{Pittau:2024ffn,
    author = "Pittau, Roberto",
    title = "{Monte Carlo evaluation of divergent one-loop integrals without contour deformation}",
    eprint = "2404.14868",
    archivePrefix = "arXiv",
    primaryClass = "hep-ph",
    doi = "10.1140/epjc/s10052-024-13109-6",
    journal = "Eur. Phys. J. C",
    volume = "84",
    number = "7",
    pages = "725",
    year = "2024"
}

@article{vanHameren:2010cp,
    author = "van Hameren, A.",
    title = "{OneLOop: For the evaluation of one-loop scalar functions}",
    eprint = "1007.4716",
    archivePrefix = "arXiv",
    primaryClass = "hep-ph",
    reportNumber = "IFJPAN-IV-2010-7",
    doi = "10.1016/j.cpc.2011.06.011",
    journal = "Comput. Phys. Commun.",
    volume = "182",
    pages = "2427--2438",
    year = "2011"
}

@article{Ellis:2007qk,
    author = "Ellis, R. Keith and Zanderighi, Giulia",
    title = "{Scalar one-loop integrals for QCD}",
    eprint = "0712.1851",
    archivePrefix = "arXiv",
    primaryClass = "hep-ph",
    reportNumber = "FERMILAB-PUB-07-633-T, OUTP-07-16P",
    doi = "10.1088/1126-6708/2008/02/002",
    journal = "JHEP",
    volume = "02",
    pages = "002",
    year = "2008"
}

@article{Soper:1999xk,
    author = "Soper, Davison E.",
    title = "{Techniques for QCD calculations by numerical integration}",
    eprint = "hep-ph/9910292",
    archivePrefix = "arXiv",
    doi = "10.1103/PhysRevD.62.014009",
    journal = "Phys. Rev. D",
    volume = "62",
    pages = "014009",
    year = "2000"
}

@article{Capatti:2019edf,
    author = "Capatti, Zeno and Hirschi, Valentin and Kermanschah, Dario and Pelloni, Andrea and Ruijl, Ben",
    title = "{Numerical Loop-Tree Duality: contour deformation and subtraction}",
    eprint = "1912.09291",
    archivePrefix = "arXiv",
    primaryClass = "hep-ph",
    doi = "10.1007/JHEP04(2020)096",
    journal = "JHEP",
    volume = "04",
    pages = "096",
    year = "2020"
}

@article{Kermanschah:2021wbk,
    author = "Kermanschah, Dario",
    title = "{Numerical integration of loop integrals through local cancellation of threshold singularities}",
    eprint = "2110.06869",
    archivePrefix = "arXiv",
    primaryClass = "hep-ph",
    doi = "10.1007/JHEP01(2022)151",
    journal = "JHEP",
    volume = "01",
    pages = "151",
    year = "2022"
}

@article{Kleiss:1994qy,
    author = "Kleiss, Ronald and Pittau, Roberto",
    title = "{Weight optimization in multichannel Monte Carlo}",
    eprint = "hep-ph/9405257",
    archivePrefix = "arXiv",
    reportNumber = "NIKHEF-H-94-17, INLO-PUB-4-94",
    doi = "10.1016/0010-4655(94)90043-4",
    journal = "Comput. Phys. Commun.",
    volume = "83",
    pages = "141--146",
    year = "1994"
}

@article{Anastasiou:2018rib,
    author = "Anastasiou, Charalampos and Sterman, George",
    title = "{Removing infrared divergences from two-loop integrals}",
    eprint = "1812.03753",
    archivePrefix = "arXiv",
    primaryClass = "hep-ph",
    doi = "10.1007/JHEP07(2019)056",
    journal = "JHEP",
    volume = "07",
    pages = "056",
    year = "2019"
}

@article{CMS,
title = {The stairway to heaven},
journal = {Physics Reports},
volume = {1115},
pages = {1-2},
year = {2025},
note = {CMS physics results from the first decade of LHC data},
issn = {0370-1573},
doi = {https://doi.org/10.1016/j.physrep.2025.01.004},
url = {https://www.sciencedirect.com/science/article/pii/S0370157325000249}
}

@article{ATLAS,
title = {Foreword to collection of ATLAS Run-2 Physics Report articles},
journal = {Physics Reports},
volume = {1116},
pages = {1-3},
year = {2025},
note = {Breaking boundaries — ATLAS physics highlights and milestones from the LHC Run 2},
issn = {0370-1573},
doi = {https://doi.org/10.1016/j.physrep.2025.01.003},
url = {https://www.sciencedirect.com/science/article/pii/S0370157325000237}
}

@article{FCC:2025lpp,
    author = "Benedikt, M. and others",
    collaboration = "FCC",
    title = "{Future Circular Collider Feasibility Study Report: Volume 1, Physics, Experiments, Detectors}",
    eprint = "2505.00272",
    archivePrefix = "arXiv",
    primaryClass = "hep-ex",
    reportNumber = "CERN-FCC-PHYS-2025-0002",
    doi = "10.1140/epjc/s10052-025-15077-x",
    journal = "Eur. Phys. J. C",
    volume = "85",
    number = "12",
    pages = "1468",
    year = "2025"
}

@article{Kotikov:1991pm,
    author = "Kotikov, A. V.",
    title = "{Differential equation method: The Calculation of N point Feynman diagrams}",
    doi = "10.1016/0370-2693(91)90536-Y",
    journal = "Phys. Lett. B",
    volume = "267",
    pages = "123--127",
    year = "1991",
    note = "[Erratum: Phys.Lett.B 295, 409--409 (1992)]"
}

@article{Gehrmann:1999as,
    author = "Gehrmann, T. and Remiddi, E.",
    title = "{Differential equations for two loop four point functions}",
    eprint = "hep-ph/9912329",
    archivePrefix = "arXiv",
    reportNumber = "TTP-99-49",
    doi = "10.1016/S0550-3213(00)00223-6",
    journal = "Nucl. Phys. B",
    volume = "580",
    pages = "485--518",
    year = "2000"
}

@article{Henn:2013pwa,
    author = "Henn, Johannes M.",
    title = "{Multiloop integrals in dimensional regularization made simple}",
    eprint = "1304.1806",
    archivePrefix = "arXiv",
    primaryClass = "hep-th",
    doi = "10.1103/PhysRevLett.110.251601",
    journal = "Phys. Rev. Lett.",
    volume = "110",
    pages = "251601",
    year = "2013"
}

@article{Caola:2014iua,
    author = "{Caola, Fabrizio and Henn, Johannes M. and Melnikov, Kirill and Smirnov, Alexander V. and Smirnov, Vladimir A.}",
    title = "{Two-loop helicity amplitudes for the production of two off-shell electroweak bosons in quark-antiquark collisions}",
    eprint = "1408.6409",
    archivePrefix = "arXiv",
    primaryClass = "hep-ph",
    reportNumber = "TTP14-026",
    doi = "10.1007/JHEP11(2014)041",
    journal = "JHEP",
    volume = "11",
    pages = "041",
    year = "2014"
}

@article{Gehrmann:2015bfy,
    author = "{Gehrmann, T. and Henn, J. M. and Lo Presti, N. A.}",
    title = "{Analytic form of the two-loop planar five-gluon all-plus-helicity amplitude in QCD}",
    eprint = "1511.05409",
    archivePrefix = "arXiv",
    primaryClass = "hep-ph",
    reportNumber = "MITP-15-101, ZU-TH-36-15",
    doi = "10.1103/PhysRevLett.116.062001",
    journal = "Phys. Rev. Lett.",
    volume = "116",
    number = "6",
    pages = "062001",
    year = "2016",
    note = "[Erratum: Phys.Rev.Lett. 116, 189903 (2016)]"
}

@article{Bonciani:2016qxi,
    author = "{Bonciani, Roberto and Del Duca, Vittorio and Frellesvig, Hjalte and Henn, Johannes M. and Moriello, Francesco and Smirnov, Vladimir A.}",
    title = "{Two-loop planar master integrals for Higgs$\to 3$ partons with full heavy-quark mass dependence}",
    eprint = "1609.06685",
    archivePrefix = "arXiv",
    primaryClass = "hep-ph",
    doi = "10.1007/JHEP12(2016)096",
    journal = "JHEP",
    volume = "12",
    pages = "096",
    year = "2016"
}

@article{Badger:2017jhb,
    author = "{Badger, Simon and Br\o{}nnum-Hansen, Christian and Hartanto, Heribertus Bayu and Peraro, Tiziano}",
    title = "{First look at two-loop five-gluon scattering in QCD}",
    eprint = "1712.02229",
    archivePrefix = "arXiv",
    primaryClass = "hep-ph",
    reportNumber = "IPPP-17-95, EDINBURGH-2017-27, MITP-17-094",
    doi = "10.1103/PhysRevLett.120.092001",
    journal = "Phys. Rev. Lett.",
    volume = "120",
    number = "9",
    pages = "092001",
    year = "2018"
}

@article{Kudashkin:2017skd,
    author = "{Kudashkin, Kirill and Melnikov, Kirill and Wever, Christopher}",
    title = "{Two-loop amplitudes for processes $g g \to H g, q g \to H q$ and $q \bar{q} \to H g$ at large Higgs transverse momentum}",
    eprint = "1712.06549",
    archivePrefix = "arXiv",
    primaryClass = "hep-ph",
    reportNumber = "TTP17-055",
    doi = "10.1007/JHEP02(2018)135",
    journal = "JHEP",
    volume = "02",
    pages = "135",
    year = "2018"
}

@article{Frellesvig:2019byn,
    author = "{Frellesvig, Hjalte and Hidding, Martijn and Maestri, Leila and Moriello, Francesco and Salvatori, Giulio}",
    title = "{The complete set of two-loop master integrals for Higgs + jet production in QCD}",
    eprint = "1911.06308",
    archivePrefix = "arXiv",
    primaryClass = "hep-ph",
    reportNumber = "MPP-2019-228",
    doi = "10.1007/JHEP06(2020)093",
    journal = "JHEP",
    volume = "06",
    pages = "093",
    year = "2020"
}

@article{Canko:2020ylt,
    author = "{Canko, Dhimiter D. and Papadopoulos, Costas G. and Syrrakos, Nikolaos}",
    title = "{Analytic representation of all planar two-loop five-point Master Integrals with one off-shell leg}",
    eprint = "2009.13917",
    archivePrefix = "arXiv",
    primaryClass = "hep-ph",
    doi = "10.1007/JHEP01(2021)199",
    journal = "JHEP",
    volume = "01",
    pages = "199",
    year = "2021"
}

@article{Agarwal:2021vdh,
    author = "{Agarwal, Bakul and Buccioni, Federico and von Manteuffel, Andreas and Tancredi, Lorenzo}",
    title = "{Two-Loop Helicity Amplitudes for Diphoton Plus Jet Production in Full Color}",
    eprint = "2105.04585",
    archivePrefix = "arXiv",
    primaryClass = "hep-ph",
    reportNumber = "MSUHEP-21-010, OUTP-21-12P",
    doi = "10.1103/PhysRevLett.127.262001",
    journal = "Phys. Rev. Lett.",
    volume = "127",
    number = "26",
    pages = "262001",
    year = "2021"
}

@article{Abreu:2021asb,
    author = "{Abreu, S. and Febres Cordero, F. and Ita, H. and Klinkert, M. and Page, B. and Sotnikov, V.}",
    title = "{Leading-color two-loop amplitudes for four partons and a W boson in QCD}",
    eprint = "2110.07541",
    archivePrefix = "arXiv",
    primaryClass = "hep-ph",
    reportNumber = "CERN-TH-2021-156, FR-PHENO-2021-12, MPP-2021-181",
    doi = "10.1007/JHEP04(2022)042",
    journal = "JHEP",
    volume = "04",
    pages = "042",
    year = "2022"
}

@article{Agarwal:2023suw,
    author = "{Agarwal, Bakul and Buccioni, Federico and Devoto, Federica and Gambuti, Giulio and von Manteuffel, Andreas and Tancredi, Lorenzo}",
    title = "{Five-parton scattering in QCD at two loops}",
    eprint = "2311.09870",
    archivePrefix = "arXiv",
    primaryClass = "hep-ph",
    reportNumber = "KA-TP-27-2023, MSUHEP-23-030, OUTP-23-12P, P3H-23-088,
  TUM-HEP-1481/23",
    doi = "10.1103/PhysRevD.109.094025",
    journal = "Phys. Rev. D",
    volume = "109",
    number = "9",
    pages = "094025",
    year = "2024"
}

@article{Abreu:2024fei,
    author = "Abreu, Samuel and Monni, Pier Francesco and Page, Ben and Usovitsch, Johann",
    title = "{Planar six-point Feynman integrals for four-dimensional gauge theories}",
    eprint = "2412.19884",
    archivePrefix = "arXiv",
    primaryClass = "hep-ph",
    reportNumber = "CERN-TH-2024-221, HU-EP-24/40-RTG",
    doi = "10.1007/JHEP06(2025)112",
    journal = "JHEP",
    volume = "06",
    pages = "112",
    year = "2025"
}

@article{Binoth:2003ak,
    author = "Binoth, T. and Heinrich, G.",
    title = "{Numerical evaluation of multiloop integrals by sector decomposition}",
    eprint = "hep-ph/0305234",
    archivePrefix = "arXiv",
    reportNumber = "EDINBURGH-2003-06, IPPP-03-28, DCPT-03-56",
    doi = "10.1016/j.nuclphysb.2003.12.023",
    journal = "Nucl. Phys. B",
    volume = "680",
    pages = "375--388",
    year = "2004"
}

@article{Bierenbaum:2010cy,
    author = "Bierenbaum, Isabella and Catani, Stefano and Draggiotis, Petros and Rodrigo, German",
    title = "{A Tree-Loop Duality Relation at Two Loops and Beyond}",
    eprint = "1007.0194",
    archivePrefix = "arXiv",
    primaryClass = "hep-ph",
    reportNumber = "IFIC-10-17",
    doi = "10.1007/JHEP10(2010)073",
    journal = "JHEP",
    volume = "10",
    pages = "073",
    year = "2010"
}

@article{Runkel:2019yrs,
    author = "Runkel, Robert and Sz\H{o}r, Zolt\'an and Vesga, Juan Pablo and Weinzierl, Stefan",
    title = "{Causality and loop-tree duality at higher loops}",
    eprint = "1902.02135",
    archivePrefix = "arXiv",
    primaryClass = "hep-ph",
    doi = "10.1103/PhysRevLett.122.111603",
    journal = "Phys. Rev. Lett.",
    volume = "122",
    number = "11",
    pages = "111603",
    year = "2019",
    note = "[Erratum: Phys.Rev.Lett. 123, 059902 (2019)]"
}

@article{Capatti:2020xjc,
    author = "Capatti, Zeno and Hirschi, Valentin and Pelloni, Andrea and Ruijl, Ben",
    title = "{Local Unitarity: a representation of differential cross-sections that is locally free of infrared singularities at any order}",
    eprint = "2010.01068",
    archivePrefix = "arXiv",
    primaryClass = "hep-ph",
    doi = "10.1007/JHEP04(2021)104",
    journal = "JHEP",
    volume = "04",
    pages = "104",
    year = "2021"
}

@article{Liu:2022chg,
    author = "{Liu, Xiao and Ma, Yan-Qing}",
    title = "{AMFlow: A Mathematica package for Feynman integrals computation via auxiliary mass flow}",
    eprint = "2201.11669",
    archivePrefix = "arXiv",
    primaryClass = "hep-ph",
    doi = "10.1016/j.cpc.2022.108565",
    journal = "Comput. Phys. Commun.",
    volume = "283",
    pages = "108565",
    year = "2023"
}

@article{Dubovyk:2022frj,
    author = "{Dubovyk, Ievgen and Freitas, Ayres and Gluza, Janusz and Grzanka, Krzysztof and Hidding, Martijn and Usovitsch, Johann}",
    title = "{Evaluation of multiloop multiscale Feynman integrals for precision physics}",
    eprint = "2201.02576",
    archivePrefix = "arXiv",
    primaryClass = "hep-ph",
    reportNumber = "CERN-TH-2021-230, UUITP-66/21",
    doi = "10.1103/PhysRevD.106.L111301",
    journal = "Phys. Rev. D",
    volume = "106",
    number = "11",
    pages = "L111301",
    year = "2022"
}

@article{Armadillo:2022ugh,
    author = "{Armadillo, Tommaso and Bonciani, Roberto and Devoto, Simone and Rana, Narayan and Vicini, Alessandro}",
    title = "{Evaluation of Feynman integrals with arbitrary complex masses via series expansions}",
    eprint = "2205.03345",
    archivePrefix = "arXiv",
    primaryClass = "hep-ph",
    doi = "10.1016/j.cpc.2022.108545",
    journal = "Comput. Phys. Commun.",
    volume = "282",
    pages = "108545",
    year = "2023"
}

@article{Borinsky:2023jdv,
    author = "{Borinsky, Michael and Munch, Henrik J. and Tellander, Felix}",
    title = "{Tropical Feynman integration in the Minkowski regime}",
    eprint = "2302.08955",
    archivePrefix = "arXiv",
    primaryClass = "hep-ph",
    reportNumber = "DESY-23-026",
    doi = "10.1016/j.cpc.2023.108874",
    journal = "Comput. Phys. Commun.",
    volume = "292",
    pages = "108874",
    year = "2023"
}

@article{Heinrich:2023til,
    author = "{Heinrich, G. and Jones, S. P. and Kerner, M. and Magerya, V. and Olsson, A. and Schlenk, J.}",
    title = "{Numerical scattering amplitudes with pySecDec}",
    eprint = "2305.19768",
    archivePrefix = "arXiv",
    primaryClass = "hep-ph",
    reportNumber = "KA-TP-09-2023, P3H-23-035, IPPP/23/24",
    doi = "10.1016/j.cpc.2023.108956",
    journal = "Comput. Phys. Commun.",
    volume = "295",
    pages = "108956",
    year = "2024"
}







\end{document}